\documentclass[useAMS]{mn2e}
\usepackage{mnras_cite}
\usepackage{epsfig}
\usepackage{deluxetable}
\usepackage{graphicx}
\begin{document}
\title[{\it Herschel} Observations of mid-IR Selected Starbursts at z$\sim$2] {Herschel reveals a $T_{\rm dust}$-unbiased selection of z$\sim$2 ULIRGs}

\author[G.E.~Magdis et al.]
{\parbox{\textwidth}{G.E.~Magdis,$^{1}$\thanks{E-mail: \texttt{georgios.magdis@cea.fr}}
D.~Elbaz,$^{1}$
H.S.~Hwang,$^{1}$
A.~Amblard,$^{3}$
V.~Arumugam,$^{4}$
H.~Aussel,$^{1}$
A.~Blain,$^{5}$
J.~Bock,$^{5,6}$
A.~Boselli,$^{7}$
V.~Buat,$^{7}$
N.~Castro-Rodr{\'\i}guez,$^{8,9}$
A.~Cava,$^{8,9}$
P.~Chanial,$^{10}$
D.L.~Clements,$^{10}$
A.~Conley,$^{11}$
L.~Conversi,$^{12}$
A.~Cooray,$^{3,5}$
C.D.~Dowell,$^{5,6}$
E.~Dwek,$^{13}$
S.~Eales,$^{14}$
D.~Farrah.$^{22}$
A.~Franceschini,$^{15}$
J.~Glenn,$^{11}$
M.~Griffin,$^{14}$
M.~Halpern,$^{16}$
E.~Hatziminaoglou,$^{17}$
J.~Huang,$^{2}$
E.~Ibar,$^{18}$
K.~Isaak,$^{14}$
R.J.~Ivison,$^{18,4}$
E.~ Le Floc'h,$^{1}$
G.~Lagache,$^{19}$
L.~Levenson,$^{5,6}$
N.~Lu,$^{5,20}$
S.~Madden,$^{1}$
B.~Maffei,$^{21}$
G.~Mainetti,$^{15}$
L.~Marchetti,$^{15}$
H.T.~Nguyen,$^{6,5}$
B.~O'Halloran,$^{10}$
S.J.~Oliver,$^{22}$
A.~Omont,$^{23}$
M.J.~Page,$^{24}$
P.~Panuzzo,$^{1}$
A.~Papageorgiou,$^{14}$
C.P.~Pearson,$^{25,26}$
I.~P{\'e}rez-Fournon,$^{8,9}$
M.~Pohlen,$^{14}$
D.~Rigopoulou,$^{40,30}$
D.~Rizzo,$^{10}$
I.G.~Roseboom,$^{22}$
M.~Rowan-Robinson,$^{10}$
B.~Schulz,$^{5,20}$
Douglas~Scott,$^{16}$
N.~Seymour,$^{24}$
D.L.~Shupe,$^{5,20}$
A.J.~Smith,$^{22}$
J.A.~Stevens,$^{27}$
M.~Symeonidis,$^{24}$
M.~Trichas,$^{10}$
K.E.~Tugwell,$^{24}$
M.~Vaccari,$^{15}$
I.~Valtchanov,$^{12}$
L.~Vigroux,$^{23}$
L.~Wang,$^{22}$
G.~Wright,$^{18}$
C.K.~Xu$^{5,20}$ and
M.~Zemcov$^{5,6}$}\vspace{0.4cm}\\
\parbox{\textwidth}{$^{1}$Laboratoire AIM-Paris-Saclay, CEA/DSM/Irfu - CNRS - Universit\'e Paris Diderot, CE-Saclay, pt courrier 131, F-91191 Gif-sur-Yvette, France\\
$^{2}$Harvard-Smithsonian Center for Astrophysics, MS65, 60 Garden Street,  Cambridge,  MA02138, USA\\
$^{3}$Dept. of Physics \& Astronomy, University of California, Irvine, CA 92697, USA\\
$^{4}$Institute for Astronomy, University of Edinburgh, Royal Observatory, Blackford Hill, Edinburgh EH9 3HJ, UK\\
$^{5}$California Institute of Technology, 1200 E. California Blvd., Pasadena, CA 91125, USA\\
$^{6}$Jet Propulsion Laboratory, 4800 Oak Grove Drive, Pasadena, CA 91109, USA\\
$^{7}$Laboratoire d'Astrophysique de Marseille, OAMP, Universit\'e Aix-marseille, CNRS, 38 rue Fr\'ed\'eric Joliot-Curie, 13388 Marseille cedex 13, France\\
$^{8}$Instituto de Astrof{\'\i}sica de Canarias (IAC), E-38200 La Laguna, Tenerife, Spain\\
$^{9}$Departamento de Astrof{\'\i}sica, Universidad de La Laguna (ULL), E-38205 La Laguna, Tenerife, Spain\\
$^{10}$Astrophysics Group, Imperial College London, Blackett Laboratory, Prince Consort Road, London SW7 2AZ, UK\\
$^{11}$Dept. of Astrophysical and Planetary Sciences, CASA 389-UCB, University of Colorado, Boulder, CO 80309, USA\\
$^{12}$Herschel Science Centre, European Space Astronomy Centre, Villanueva de la Ca\~nada, 28691 Madrid, Spain\\
$^{13}$Observational  Cosmology Lab, Code 665, NASA Goddard Space Flight  Center, Greenbelt, MD 20771, USA\\
$^{14}$Cardiff School of Physics and Astronomy, Cardiff University, Queens Buildings, The Parade, Cardiff CF24 3AA, UK\\
$^{15}$Dipartimento di Astronomia, Universit\`{a} di Padova, vicolo Osservatorio, 3, 35122 Padova, Italy\\
$^{16}$Department of Physics \& Astronomy, University of British Columbia, 6224 Agricultural Road, Vancouver, BC V6T~1Z1, Canada\\
$^{17}$ESO, Karl-Schwarzschild-Str. 2, 85748 Garching bei M\"unchen, Germany\\
$^{18}$UK Astronomy Technology Centre, Royal Observatory, Blackford Hill, Edinburgh EH9 3HJ, UK\\
$^{19}$Institut d'Astrophysique Spatiale (IAS), b\^atiment 121, Universit\'e Paris-Sud 11 and CNRS (UMR 8617), 91405 Orsay, France\\
$^{20}$Infrared Processing and Analysis Center, MS 100-22, California Institute of Technology, JPL, Pasadena, CA 91125, USA\\
$^{21}$School of Physics and Astronomy, The University of Manchester, Alan Turing Building, Oxford Road, Manchester M13 9PL, UK\\
$^{22}$Astronomy Centre, Dept. of Physics \& Astronomy, University of Sussex, Brighton BN1 9QH, UK\\
$^{23}$Institut d'Astrophysique de Paris, UMR 7095, CNRS, UPMC Univ. Paris 06, 98bis boulevard Arago, F-75014 Paris, France\\
$^{24}$Mullard Space Science Laboratory, University College London, Holmbury St. Mary, Dorking, Surrey RH5 6NT, UK\\
$^{25}$Space Science \& Technology Department, Rutherford Appleton Laboratory, Chilton, Didcot, Oxfordshire OX11 0QX, UK\\
$^{26}$Institute for Space Imaging Science, University of Lethbridge, Lethbridge, Alberta, T1K 3M4, Canada\\
$^{27}$Centre for Astrophysics Research, University of Hertfordshire, College Lane, Hatfield, Hertfordshire AL10 9AB, UK}}

\maketitle

\clearpage
\begin{abstract}
Using {\it Herschel} PACS and SPIRE observations of Lockman Hole-North and GOODS-N as part of the HerMES project, we explore the far-IR properties of a sample of mid-IR selected starburst dominated ultra-luminous infrared galaxies (ULIRGs) at z $\sim$ 2. The selection of the sample is based on the detection of the stellar bump that appears in the SED of star-forming galaxies at 1.6$\mu$m. We derive robust estimates of infrared luminosities ($L_{\rm IR}$) and dust temperatures ($T_{\rm d}$) of the population and find that while the luminosities in our sample  span less than an order of magnitude ($12.24\leq log(L_{\rm IR}/L_{\odot}) \leq 12.94$), they cover a wide range of dust temperatures ($25\leq T_{\rm d} \leq 62$ K). Galaxies in our sample range from those that are as cold as high-z sub-millimeter galaxies (SMGs) to those that are as warm as optically faint radio galaxies (OFRGs) and local ULIRGs. Nevertheless, our sample has median $T_{\rm d}$=42.3 K, filling the gap between SMGs and OFRGs, bridging the two populations. We demonstrate that a significant fraction of our sample would be missed from ground based (sub)mm surveys (850-1200$\mu$m) showing that the latter introduce a bias towards the detection of colder sources. We conclude that {\it Herschel} observations, confirm the existence of high-z ULIRGs warmer than SMGs, show that the mid-IR selection of high-z ULIRGs is not $T_{\rm d}$-dependent, reveal a large dispersion in $T_{\rm d}$ of high-z ULIRGs, and provide the means to characterize the bulk of the ULIRG population, free from selection biases introduced by ground based (sub)mm surveys.

\end{abstract}

\section{Introduction}
One of the most successful methods for selecting high-z ultra-luminous infrared galaxies (ULIRGs: $L_{\rm 8-1000}$ $_{\mu m}$  $>$ 10$^{12}$ $L_{\odot}$) is their direct far-IR detection via ground based (sub)millimeter surveys (e.g. Barger et al. 1998, Hughes et al. 1998, Mortier et al. 2005, Pope et al. 2006, Austermann et al. 2010). This technique has revealed the population of the so called submillimetre galaxies (SMGs), that represent a significant class of high-z ULIRGs. Attempts to characterize their dust temperature ($T_{\rm d}$) show that these galaxies are colder when compared to local ULIRGs  (e.g Chapman et al. 2005), suggesting that in general high-z ULIRGs tend to have lower dust temperatures. However, the submillimetre technique introduces a bias towards the selection of ULIRGs with lower dust temperatures while it misses warmer ULIRGs. First observational evidence of a missing population of high-redshift dusty star-forming
galaxies with hotter dust has been given by Chapman et al. (2004)
using a selection of radio-detected but sub-mm-faint galaxies
with UV spectra consistent with high-redshift starbursts. These
optically faint-radio galaxies (OFRGs) share similar properties with SMGs (e.g. stellar mass, SFR)  but some have considerably higher dust temperatures (Casey et al. 2009, Magnelli et al. 2010).

Another technique that has been proven to pick high-z starburst 
dominated ULIRGs efficiently, is based on mid-IR color selection. This technique relies on the detection of the rest-frame 1.6$\mu$m bump in the SED of star-forming galaxies, produced by thermal emission from late-type stars and enhanced by an apparent emission feature due to H- ions in the atmospheres of giant stars (Simpson \& Eisenhardt 1999; Sawicki 2002). The advent of {\it Spitzer} allowed the detection of this feature in z$\sim$2 galaxies and subsequent IRS spectroscopy has demonstrated the efficiency of the method to select star-burst dominated ULIRGs in a redshift range of 1.5 $<$ z $<$ 2.5 (e.g. Farrah et al 2008, Weedman et al 2008, Huang et al. 2009). Further studies (Lonsdale et al. 2009, Fiolet et al. 2009, Kovacs et al. 2010), indicate that only 40$\%$ of their sample is made up of bright mm sources and thus belong to the class of SMGs, while most of the rest have lower $S_{1.2mm}$ fluxes.

The above suggests that a considerable fraction of mid-IR selected high-z ULIRGs are missed by ground based (sub)-mm surveys. While this could naturally be explained if their dust temperature (for a given luminosity) is higher than that of the SMGs, this is not yet clear, as up to now the study of their far-IR properties is restricted to objects with ground (sub)mm detection or to the most luminous examples of the population with the highly confused BLAST beam (e.g Ivison et al. 2010, Dunlop et al. 2009). Hence, a far-IR study of the population, free of the selection bias introduced by the ground based sub-mm  detection is required. Furthermore, detailed study of high-z ULIRGs is essential as up to now theoretical models fail to account for the inferred luminosities, star formation rates and number counts (Baugh et al. 2005, Dave et al. 2010).

In this study, we use observations of Lockman-North (LHN) and GOODS-N fields obtained by the {\it Herschel} Space Observatory (Pilbratt et al. 2010) as part of the Herschel Multi-tiered Extragalactic Survey (HerMES, Olivier et al. 2010), to investigate the FIR properties of a sample of mid-IR selected ULIRGs at z$\sim$2 (IRAC peakers). Taking advantage of both the Photodetector Array Camera (PACS, Poglisch et al. 2010) and the Spectral and Photometric Imaging Receiver (SPIRE, Griffin et al. 2010) data that probe the peak of the SED of galaxies at this redshift, we derive robust dust temperature measurements for the bulk of the population and compare our sample to that of other high-z ULIRGs. Throughout this paper we assume $\Omega_{\rm m}$=0.3, H$_{0}$=71km sec$^{-1}$ Mpc$^{-1}$ and $\Omega_{\Lambda}$=0.7       

\begin{figure*}
\centering
\includegraphics[width=0.45\textwidth,angle=0]{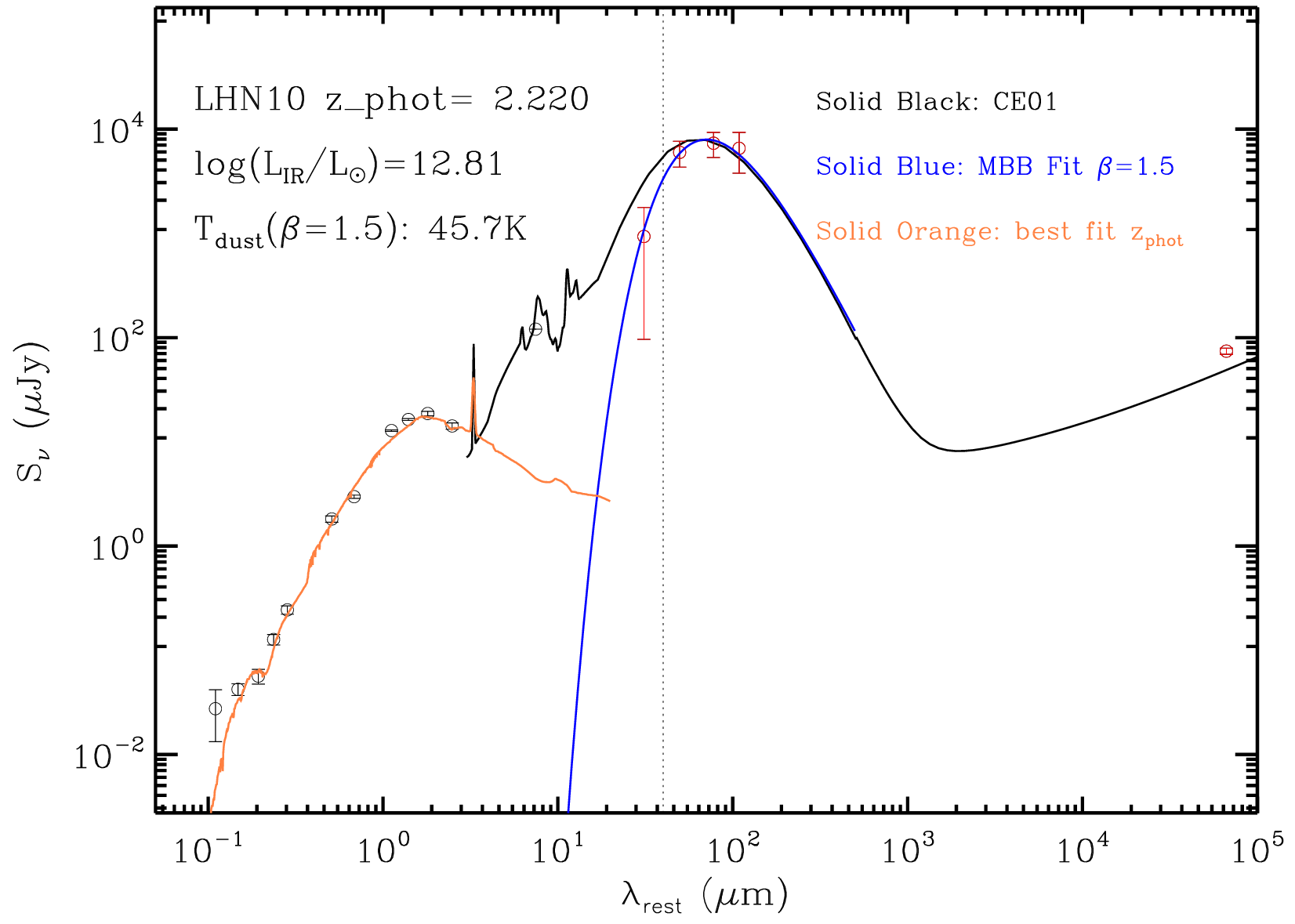}\hspace*{0.01\textwidth}
\includegraphics[width=0.45\textwidth,angle=0]{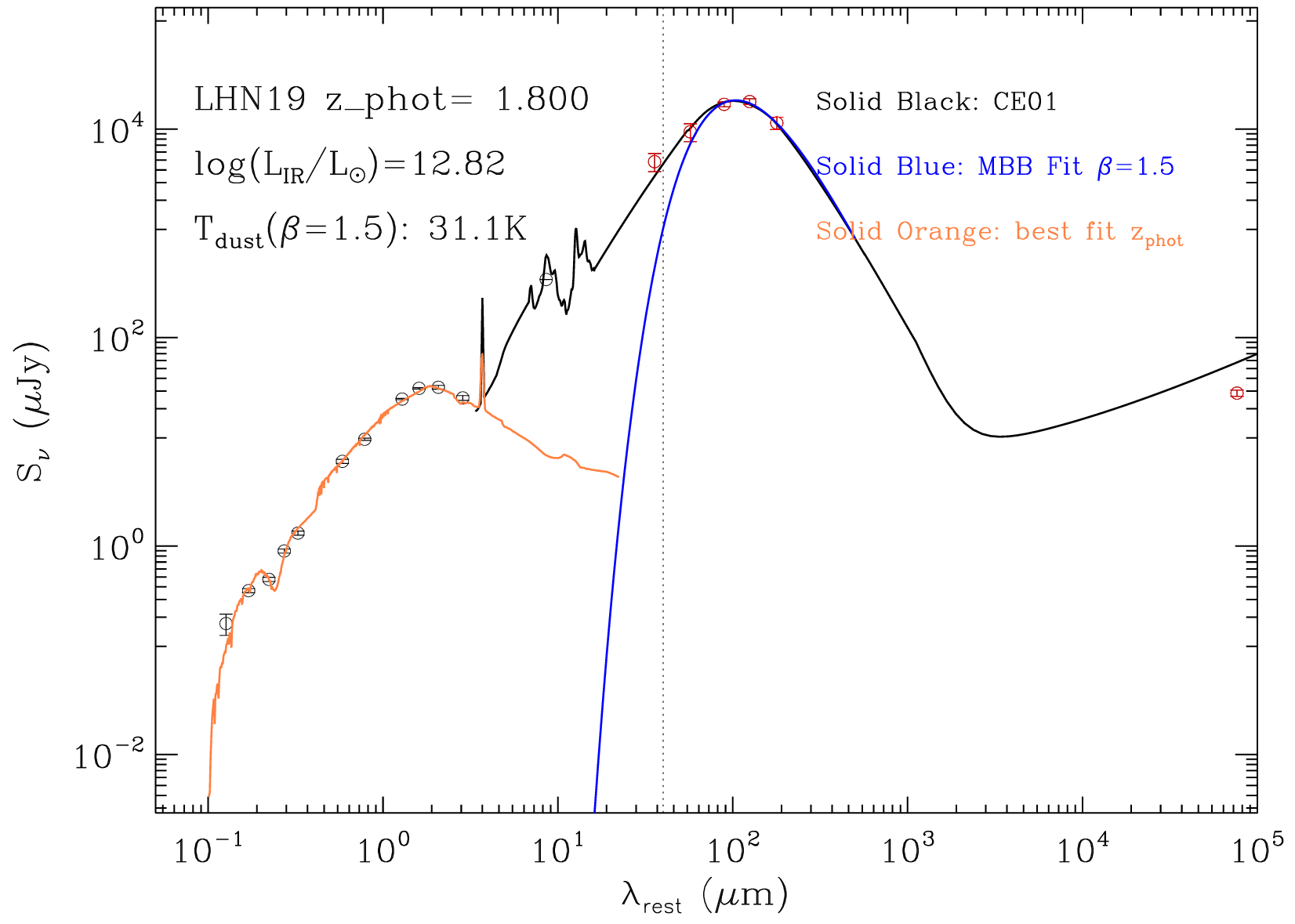}\hspace*{0.01\textwidth}
\caption{Rest--frame SEDs and derivation of the far-IR properties for two ULIRGs in our sample. Solid orange line shows the best fit template up to observed 8$\mu$m as derived by LEPHARE photo-z code. Solid black line shows the best fit CE01 model while the blue line depicts the best-fit modified black body (with $\beta$=1.5), used to derive $T_{\rm d}$ estimates. The vertical dotted line indicates the wavelength cut, below which photometric data where not considered in the modified black body fit. Red circles denote {it Herschel} data.}
\label{fig:sub} %
\end{figure*}
\section{Sample Selection and {\it Herschel} Observations}
To select our sample we adopt the IRAC color criteria introduced by Huang e al. (2009). In particular we search for galaxies in LHN that satisfy the following IRAC color criteria: 0.05$<$[3.6]$-$[4.5]$<$0.4 and $-$0.7$<$[3.6]$-$[8.0]$<$0.5,  have f$_{24}$$>$0.2mJy and r$_{\rm vega}$$>$23.0 to avoid low redshift interlopers. These criteria probe the 1.6$\mu$m stellar bump while the red color cuts ensure the rejection of power-law AGNs. Subsequent IRS spectroscopy of ULIRGs selected with the above method, has shown that this selection picks ULIRGs at very narrow redshift range 1.7$<$z$<$2.3 with strong PAH features, indicative of intensive star-formation (Huang et al. 2009).

We use the Spitzer Wide-Area Infrared Extragalactic Survey (SWIRE) multi-wavelength catalog (U,G,R,I,z,J,H,K+IRAC+MIPS) (Surace et al. 2005) over the 0.25(sq deg) of LHN covered by PACS and SPIRE, and we identify 32 objects that meet our criteria. We then match the sample with the 
joined {\it Herschel} PACS (100- and 160$\mu$m and SPIRE (250- 350- and 500$\mu$m) XID catalog  (Roseboom et al. 2010). For the PACS data, where source confusion is less severe, we consider fluxes based on blind source extraction. For the SPIRE data we adopt the fluxes derived with source extraction based on 24$\mu$m 
priors, we reject candidates with neighboring (d$<$20'') 24$\mu$m 
sources whose $f_{24}$ is $>$ 50\% than that of our object. Finally, we require at least one PACS and SPIRE detection (3$\sigma$). 

In the resulting LHN sample there are 25 candidate ULIRGs at z$\sim$2, five of which have spectroscopic redshifts (Fiolet et al. 2010, in prep). For the rest, we derive photometric redshifts, using the LEPHARE photo-z code (Ilbert et al. 2009). Namely, we fit the SED of the galaxies up to  8.0$\mu$m with a wide range of template SEDs and consider dust attenuation that follows the prescription of Caltzetti et al. (2000). For each object we adopt the redshift corresponding to the minimum $\chi^{2}$ value of the fit. Two examples of the best fit template for two galaxies in our sample are depicted in Figure 1. The uncertainty of the photometric redshifts was derived based on the redshift probability distribution function ( PDF(z) ) and we choose to exclude from our analysis candidates with multiple solutions or 
uncertainties larger than $\Delta z$ = 0.5. Furthermore, a comparison of the derived photometric redshifts with the spectroscopic redshifts that is available for 5 sources, yields a very good agreement between the two values $\Delta$$z$=(z$_{\rm photo}$$-$z$_{\rm spec}$)/(1+z$_{\rm spec}$) $<$ 0.1. The final LHN sample consists of 18 ULIRGs with median z=1.98, and range 1.5$<$z$<$3.0. Finally, we match this sample to radio VLA 1.4GHz catalog of LHN (Owen et al. 2008).

To increase the size of our sample, we perform the same procedure in the GOODS-N field. Using the multi-wavelength catalog and the SPIRE data (PACS data for GOODS-N are not available for this study), we indentify candidate ULIRGs with at least two detections (3$\sigma)$ at SPIRE bands. To exclude sources with strong AGN activity
candidates with X-ray detection ($L_{\rm X}$[0.5-8.0keV] $>$ 3$\times$10$^{42}$ ergs s$^{-1}$)  were removed. The final GOODS-N sample consists of seven sources. Out of these, one has spectroscopic redshift (z=1.86) while for the rest we adopt photometric redshift by Le Borgne et al. 2009 (median z=1.83 and 1.53$<$z$<$2.05). The final combined sample (LHN and GOODS-N) consists of 25 ULIRGs with a median z=2.01 and with 18 out of 25 objects lying in narrow redshift range (1.7$<$z$<$2.3). PACS/SPIRE photometry of our sample is presented in Table 1 (electronic version), while IRAC, MIPS and mm and radio photometry is given in Table 2 (electronic version) 

\section{Derivation of far-IR properties}
To derive estimates for the $L_{\rm IR}$ ($L_{\rm 8-1000\mu m}$) of the galaxies in our sample, we first convert their SED to rest-frame applying k--corrections and then fit the PACS and SPIRE data with the libraries of Chary \& Elbaz (2001) (CE01) and Dale \& Helou (2002). Results based on the two methods are in very close agreement indicating a median $L_{\rm IR}$=3$\times$10$^{12}$ $L_{\odot}$. The CE01-derived $L_{\rm IR}$ for each object are summarized in Table 1, while examples of the rest--frame SEDs along with the best-fit CE01 templates for two ULIRGs in our sample are shown in Figure 1.

To derive the dust temperature of galaxies in our sample, we use a single temperature modified black body fitting form in which the thermal dust spectrum is approximated as $F_{\nu} \propto \nu^{3+\beta}/($e$^{(h\nu/kT_{d})}-1)$. This model was fit to {\it Herschel} data with rest--frame $\lambda$ $>$ 40$\mu$m, assuming a fixed emmisivity index of $\beta$=1.5. This wavelength cut was introduced to avoid fitting emission from Very Small Grains (VSGs). The $T_{\rm d}$ of each object was obtained from the best fit model, based on the minimization of the $\chi^{2}$ value. The uncertainty for each $T_{\rm d}$ value was estimated by repeating the same procedure for random perturbations of the fitted photometric points within their errors (following a normal distribution). The best fit model for two ULIRGs in our sample are shown in Figure 1 (solid blue line) with the $T_{\rm d}$ for each of the galaxies summarized in Table 1. Finally, to check whether the lack of PACS data for the GOODS-N sample introduces any systematic bias in the derived properties we repeated the fitting procedure for the LHN sample, excluding this time the PACS photometric points. The values derived with and without the PACS were in good agreement ($<$ $\Delta$ $T_{\rm d}$ $>$ = 1.9 K). 

\subsection{AGN contribution to our sample}
It has been shown by previous studies that the selection criteria of our sample have been very successful in selecting starburst over AGN dominated ULIRGs, (e.g. Farrah et al. 2008, Huang et al. 2009). Indeed, for five ULIRGs in LHN, IRS spectroscopy indicates that their mid-IR emission is dominated by vigorous star-formation rather than an AGN (Fiolet et al. 2010 in prep). None of our objects in this field is detected by Chandra to a 0.3-2.5keV flux limit of $5\times 10^{-16}$ erg cm$^{-2}$s$^{-1}$ s(L$_{x}$ $>$8.5$\times$10$^{42}$ erg s$^{-1}$ for z=2.0) (Polletta et al. 2006), while on construction of the GOODS-N sample all candidate objects with X-ray detection at a flux limit of $1.95\times 10^{-17}$ erg cm$^{-2}$s$^{-1}$ were rejected from our analysis. As the moderate depth of the LHN X-ray data do not provide strong constraints on the AGN contribution we further explore this issue by the $q$ parameter ($q=log [L_{\rm 40-120\mu m} / (3.75 \times 10^{12} \rm ~W]$  $- log [L_{\rm 1.4GHz}/(\rm W~Hz^{-1})]$, Helou 1985). Given that almost all of our galaxies have radio detections, we estimate the $q$ parameter and find a mean $<q>$=2.21 witssh intrinsic dispersion $\sigma_{q}$=0.17. This is in agreement with that found by Younger et al. (2009) and the $q$ of star-forming galaxies, quoted by Ivison et al (2010) ($q=2.40$, 2$\sigma_{q}$=0.27). These considerations, support the conclusion of Huang et al. (2009) that an AGN contributes little ($<10-20\%$) to the bolometric luminosity of these objects.  
 
\section{Results}
\subsection{Far-IR properties and comparison with other ULIRG samples}
Galaxies in our sample have dust temperatures that span a wide range $25\leq T_{\rm d} \leq 62$ (K), while their luminosities vary by less than an order of magnitude $12.24\leq log(L_{\rm IR}/L_{\odot}) \leq 12.94$. The median values are $T_{\rm d}$=42.3 K, and $L_{\rm IR}$=3$\times$10$^{12}$ $L_{\odot}$, indicating a star formation rate of $\sim$520 M$_{\odot}$ yr$^{-1}$ (assuming Salpeter IMF). It is interesting to compare these values to that of ULIRG samples selected by different techniques.

We consider a large set of z$\sim$2 SMGs (Chapman et al. 2005 and Kovacs et al. 2006), a sample of z$\sim$2 OFRGs (Casey et al. 2009) and a compilation of local/intermediate-z (0$<$z$<$0.98) ULIRGs (Clements et al. 2010, Farrah et al. 2003 and Yang et al. 2007). In all these studies, the method to derive $T_{\rm d}$ estimates is similar to ours, fitting modified black-body models to the far-IR photometric points. For studies that quote $L_{\rm FIR}$ instead of $L_{\rm IR}$ we adopt the following conversion factor between the two values: $L_{\rm IR}$ = 1.19$\times$ $L_{\rm FIR}$ (Dale et al. 2001). We note that preliminary results by Hwang et al. 2010 (in prep) and Chapman et al. 2010 (in prep) and Chanial et al. 2010 (in prep) indicate that the far-IR properties of SMGs and OFRGs when {\it Herschel} data are taken into account, are, in general, consistent with the results obtained in the pre-{\it Herschel} era.
   \begin{figure}
   \centering
   \includegraphics[width=8cm,height=8cm]{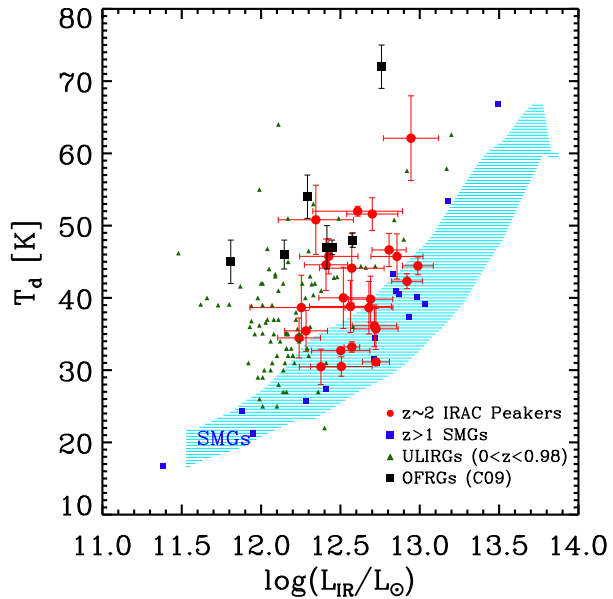}
   \caption{The $L_{\rm IR}-T_{\rm d}$ relation for our sample (red circles). Included are results for local/intermediate-z ULIRGs (green filled triangles, Farrah et al. 2003, Clements et al. 2010, Yang et al. 2007), high-z SMGs (blue squares, Chapman et al. 2005, Kovacs et al. 2006) and OFRGs (black squares, Casey et al. 2009). The cyan shaded area denotes the 2$\sigma$ envelope of the $L_{\rm IR}$-$T_{\rm d}$ relation of high-z SMGs. For a given L$_{\rm IR}$, our sample span in a wide range of dust temperatures, bridging the ``cold" high-z SMGs to the ``warmer'' local/intermediate-z ULIRGs and z$\sim$2 OFRGs. }
    \label{FigGam}%
    \end{figure}   

In Figure 2 we show the $L_{\rm IR}-T_{\rm d}$ relation for our sample as compared to that of local/intermediate-z ULIRGs, SMGs and OFRGs. For the luminosity bin of our sample, SMGs have a median $T_{\rm d}$ = 36 $\pm$ 8 K while OFRGs are considerably warmer with median $T_{\rm d}$=47 $\pm$ 3 K (Magnelli et al 2010) and dust temperatures similar to that of local ULIRGs. Therefore, it appears that the two methods select ULIRGs with significantly different dust temperatures, and with no significant overlap between them. Taking advantage of the wealth of multi-wavelength data in GOODS-N, we find that a large fraction of SMGs (15/24) and OFRGs (4/5) in GOODS-N (Pope et al. 2006, Casey et al. 2009) that fall in the redshift bin (1.5$<$z$<$3.0) of our sample, satisfy the IRAC-peakers colour criteria, if we relax the $f_{\rm 24}$ cut.

Based on this plot there are a number of significant results to be drawn. First of all, 
our observations confirm the existence of ULIRGs in the high-z universe with dust temperature 
higher than that of SMGs. Furthermore, it seems that the selection of high-z ULIRGs based on the detection of the 1.6$\mu$m bump doen't favour a particular $T_{\rm d}$, selecting ULIRGs that overlap with the SMGs and OFRGs but also ULIRGs of intermediate $T_{\rm d}$. Indeed, we see that objects in our sample range from those that are as cold as SMGs to objects as warm as OFRGs, while a significant fraction lies in the intermediate region between the two samples, bridging the two populations. We also note that a large fraction of the sample falls in the $T_{\rm d}$-$L_{\rm IR}$ relation of the local ULIRGs. Finally, our data indicate that the $T_{\rm d}$ dispersion of high-z ULIRGs is larger than that of the local ULIRGs as derived based on IRAS observations. This discrepancy mainly arises due to absence of cold sources in the local universe, although the IRAS selection might miss existing cold sources, introducing a bias towards warmer ULIRGs.  


\subsection{SMGs: Evidence of selection bias towards colder ULIRGs}
There is growing evidence that ground based (sub)mm observations introduce a systematic bias towards the detection of cold ULIRGs. As mentioned above this was first discussed by Chapman et al. (2004), introducing the populations of OFRGs while a similar conclusion was reached recently by Chapin et al. (2010) using BLAST data. In Figure 2 we showed that a fraction of IRAC peakers also tends to be warmer than high-z SMGs. We now ask whether these IRAC-peakers would be missed by the sub-mm selection.

To investigate this, we estimate the $S_{\rm 850}$ flux densities of our sample based on the best fit CE01 model that was obtained through the fitting of the {\it Herschel} photometric points. The predicted $S_{850}$ fluxes of our sample along with the measured sub-mm flux of high-z SMGs are plotted over the derived $T_{\rm d}$ of the two populations in figure 3. We also overplot tracks in constant $L_{\rm IR}$. This plot illustrates that a significant fraction ($\sim$60\%) of the mid-IR selected ULIRGs in our sample have $S_{\rm 850}$ flux densities lower than that of the SMGs, lie below the confusion limit at 850$\mu$m (2-3 mJy, Knudsen et al. 2008) and hence would be missed by ground-based (sub)mm surveys. Nevertheless, we also find IRAC-peakers with predicted $S_{\rm 850}$ above the detection limit and which therefore should be detected in the sub-mm. Indeed, four of our objects in LNH (LHN1, LHN8, LHN16, LHN29), have been detected ($S/N > 3$) by MAMBO 1.2mm (Fiolet et al. 2009, Kovacs et al. 2010). For these objects we use the formula described by Ivison et al. (2005) to convert the observed 1.2mm to 850$\mu$m flux densities and then compare these values with the predicted $S_{850}$ flux densities that we derived from our analysis. The two values are in close agreement for all objects, with a median difference of 0.15mJy. Furthermore, we find that all galaxies in our sample with MAMBO observations but no detection (LHN0,LHN19,LHN25, Fiolet et al. 2009) have predicted fluxes below the detection limit. The same test for the GOODS-N sample reveals that our analysis, successfully predicts the sub-mm fluxes of two objects with SCUBA 850$\mu$m detection (GN17, GN06, Borys et al. 2005, Pope et al. 2006). 

\begin{figure}
   \centering
   \includegraphics[width=8cm,height=8cm]{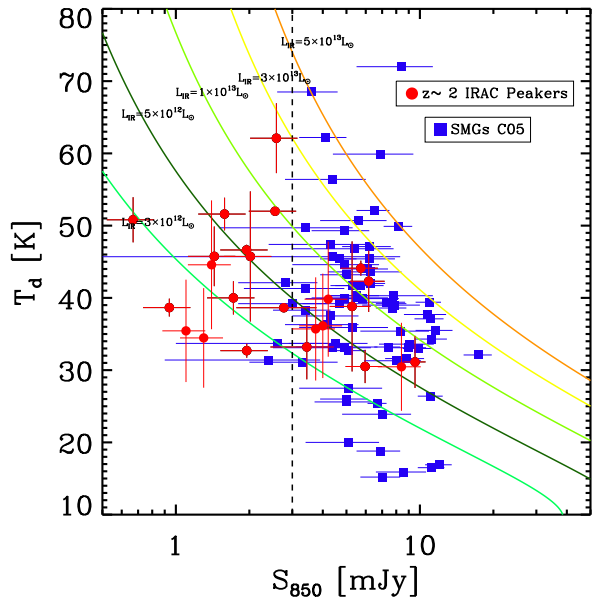}
   \caption{Dust temperature versus the estimated $S_{\rm 850}$ flux densities of galaxies in our sample (red circles). We also include $T_{\rm d}$ measurements and observed $S_{\rm 850}$ flux densities of high-z SMGs by Chapman et al, (2005) (blue squares). Solid lines represent tracks in constant $L_{\rm IR}$ while the vertical dotted line indicates the confusion limit of current ground based submm surveys. It is evident that a significant fraction of our sample lies below the detection limit and would be missed the SCUBA-850$\mu$m surveys, if we consider that the detection limit should be above the confusion.}
    \label{FigGam}%
    \end{figure}

bf Another way to explore this issue, is a direct visualization of the SEDs of the sources. Such approach is free of systematics and uncertainties of SED fitting, 
that could possibly affect/bias our results. In Figure 4, we show the far-IR SED of four sources in LH-N that have been followed up by MAMBO 1.2mm observations. Two sources are not detected at 1.2mm while the other two have a $>$ 3$\sigma$ detection. These sources are also chosen to have sililar $L_{\rm IR}$ ( 12.5 $<$ log ($L_{\rm IR}$/$L_{\odot}$) $<$ 12.7) and silimar fluxes at 250$\mu$m. It is evident that for the same luminosity, the far-IR SED of the MAMBO-undetected sources, peaks at shorter wavelengths compared to that of the MAMBO detected sources. 
This indicates a clear difference of the dust temperature and subsequently of the 850-1200$\mu$m emission of the two samples, with MAMBO-undetected sources having warmer $T_{\rm d}$ and considerably lower 850-1200$\mu$m flux densities.


Recently, Kovacs et al. (2010), presented a far-IR study of 20 luminous z$\sim$2 mid-IR selected starbursts 
based on SHARC-2 350$\mu$m and concluded that their properties are indistinguishable from the purely SMGs population. Although this seems to contradict our findings this is not the case. Since their study focuses on IRAC-peakers with S$_{\rm 1.2mm}$ $>$ 2mJy, their sample is biased towards the most sub-mm luminous galaxies among the mid-IR selected ULIRGs and hence those that are likely to share similar properties with the SMGs. As illustrated in Figure 3 such galaxies exist in our sample too. In fact, our sample shares four objects in common with that of Kovacs et al. (2010), for which the estimates of the far-IR properties between the two studies are in very good agreement. On the other hand, as we have shown above, due to the requirement of MAMBO detection, they miss a large fraction of mid-IR selected ULIRGs that have faint 850-1200$\mu$m flux densities and their properties are different from that of SMGs. Furthermore, although there are no 850-1200$\mu$m observations for some galaxies in our sample, the fraction of galaxies with predicted $S_{850}$ above the detection limit is consistent with the fraction of MAMBO detected mid-IR ULIRGs ($\sim$40\%) in the study of Lonsdale et al. (2009) and Fiolet et al. (2009). To summarise, {\it Herschel} data allow us for the first time to characterize the far-IR properties of $\sim$~50$\%$ of the mid-IR selected ULIRGs that would be missed by  ground based (sub)mm surveys and reveal that their properties are different from that of SCUBA/IRAM selected galaxies. 
\begin{figure}
   \centering
   \includegraphics[width=8cm,height=7cm]{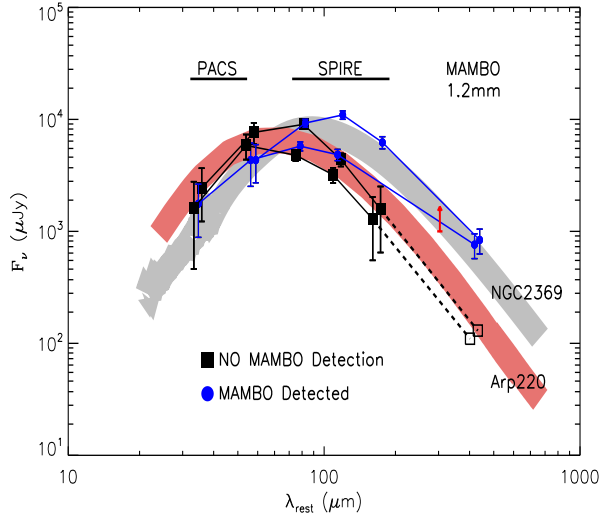}
   \caption{The far-IR part of the SED of two MAMBO-detected (blue circles) and two MAMBO-undetected sources (black squares) from our sample. 
   All four sources are chosen to
   have comparable $L_{\rm IR}$ and $f_{\rm 250}$. The far-IR part of the SED of Arp220 (coral shaded area) and that of NGC2369 (grey shaded area) 
   are also shown. For the MAMBO undetected soures the open boxes correspond to the 1.2mm flux density based on the SED extrapolation. 
  The red arrow indicates the confusion limit of 850$\mu$m surveys. The SED of MAMBO-undetected sources peaks at shorter wavelenghts, 
  indicating warmer $T_{\rm d}$ and lower 850-1200$\mu$m emission when compared to that of IRAM detected sources. This plot illustrates that 
   among sources with comparable $L_{\rm IR}$, those with higher $T_{\rm d}$ are missed by current ground-based surveys.}
    \label{FigGam}%
    \end{figure}

\section{Conclusions}
We have presented the far-IR properties of mid-IR selected ULIRGs at z$\sim$2 in LHN and GOODS-N fields, based on {\it Herschel} PACS and SPIRE observations. We showed that for a narrow range of luminosities, our sample spans a wide range of $T_{\rm d}$, indicating that the mid-IR selection of high-z ULIRGs doesn't introduce a systematic bias in $T_{\rm d}$. Sources in our sample range from those that are as cold as high-z SMGs to objects as warm as OFRGs, while a significant fraction has intermediate $T_{\rm d}$, bridging the two populations. We also demonstrated that a significant fraction of our sample would be missed from (sub)mm surveys, showing that the sub-mm technique introduces a bias towards the detection of colder ULIRG sources. We
  confirmed the existence of star-forming ULIRGs at high-z that are warmer than SMGs and showed that the $T_{\rm d}$ dispersion at high-z is larger than that found in the local universe. While this large dispersion in $T_{\rm d}$ suggests a diversity of the physical mechanisms that drive the star-formation activity in the early galaxies, its origin remains unclear. {\it Herschel} observations of larger samples in the rest of the HerMES survey, will address this question as well as the contribution of ULIRGs to the star-formation density and their clustering properties.


\section{Acknowledgments} 
SPIRE has been developed by a consortium of institutes led by Cardiff Univ. (UK) and including Univ. Lethbridge (Canada); NAOC (China); CEA, LAM (France); IFSI, Univ. Padua (Italy); IAC (Spain); Stockholm Observatory (Sweden); Imperial College London, RAL, UCL-MSSL, UKATC, Univ. Sussex (UK); Caltech, JPL, NHSC, Univ. Colorado (USA). This development has been supported by national fund- ing agencies: CSA (Canada); NAOC (China); CEA, CNES, CNRS (France); ASI (Italy); MCINN (Spain); SNSB (Sweden); STFC (UK); and NASA (USA). HIPE is a joint development (are joint developments) by the {\it Herschel} Science Ground Segment Consortium, consisting of ESA, the NASA {\it Herschel} Science Center, and the HIFI, PACS and SPIRE consortia.” The data presented in this paper will be released through the {\it Herschel} Database in Marseille HeDaM (hedam.oamp.fr/HerMES).

\begin{table*}
\centering
\caption{Far-IR properties of mid-IR selected z$\sim$2 ULIRGs.}
\label{catalog}

\renewcommand{\footnoterule}{}  
\begin{tabular}{lcccccccc}
\hline \hline
ID& {\it z} & {\it S$_{100}$$^{b}$} & {\it S$_{160}$$^{b}$} & {\it S$_{250}$} & {\it S$_{350}$} & {\it S$_{500}$}& L$\rm_{IR}$ & Td \\
 & &[{\rm mJy}] & [{\rm mJy}] & [{\rm mJy}] & [{\rm mJy}] & [{\rm mJy}] & [{\rm L$_{\odot}$}] & [{\rm K}] \\
\hline
 LHN0   &   2.01$^{a}$  &   7.36  $\pm$   3.67&  23.04  $\pm$   4.99&  27.17  $\pm$	1.94&  13.20  $\pm$   1.77&   4.76  $\pm$   2.81&  12.70  $\pm$  0.16  &  51.61  $\pm$  2.26  \\
 LHN1   &   1.95$^{a}$  &  13.37  $\pm$   2.50&  41.78  $\pm$   5.96&  52.51  $\pm$	2.14&  34.21  $\pm$   2.86&   5.02  $\pm$   5.67&  12.99  $\pm$  0.10  &  44.43  $\pm$  1.24  \\
 LHN2   &   2.01  &   5.12  $\pm$   1.68&   0.00  $\pm$   0.00&   7.36  $\pm$	1.57&	6.01  $\pm$   1.74&   0.53  $\pm$   1.97&  12.25  $\pm$  0.32  &  38.66  $\pm$  4.47  \\
 LHN3   &   2.40  &   8.94  $\pm$   2.80&   0.00  $\pm$   0.00&  15.11  $\pm$	1.71&  15.20  $\pm$   2.12&  14.93  $\pm$   2.38&  12.57  $\pm$  0.05  &  33.20  $\pm$  0.72  \\
 LHN4   &   1.72  &  20.05  $\pm$   2.83&  35.52  $\pm$   6.37&  32.53  $\pm$	7.87&  10.83  $\pm$  15.04&   0.00  $\pm$  0.00&  12.81  $\pm$  0.11  &  46.64  $\pm$  2.30  \\
 LHN5   &   1.98  &   8.70  $\pm$   2.59&   0.00  $\pm$   0.00&  16.86  $\pm$	1.64&  13.12  $\pm$   1.95&   7.07  $\pm$   2.23&  12.52  $\pm$  0.15  &  40.00  $\pm$  4.25  \\
 LHN8   &   2.07$^{a}$  &  11.07  $\pm$   2.53&  46.10  $\pm$   5.23&  46.65  $\pm$	2.03&  41.73  $\pm$   1.67&  17.70  $\pm$   2.19&  12.92  $\pm$  0.10  &  42.30  $\pm$  1.04  \\
 LHN10   &   2.22  &   3.01  $\pm$   2.70&  19.29  $\pm$   5.36&  23.60  $\pm$   6.40&  21.09  $\pm$   8.92&   0.00  $\pm$  0.00&  12.86  $\pm$  0.16  &  45.74  $\pm$  3.12  \\
 LHN11   &   2.09  &   6.18  $\pm$   2.01&   0.00  $\pm$   0.00&  13.87  $\pm$   2.07&  14.70  $\pm$   2.40&   0.00  $\pm$  0.00&  12.50  $\pm$  0.18  &  32.70  $\pm$  0.36  \\
 LHN16   &   2.10$^{a}$  &   5.47  $\pm$   2.74&  13.44  $\pm$   4.41&  29.35  $\pm$   6.31&   37.53  $\pm$  8.47&   0.00  $\pm$  0.00&  12.57  $\pm$  0.20  &  44.11  $\pm$  4.83  \\
 LHN19   &   1.80  &  13.65  $\pm$   2.69&  26.41  $\pm$   5.18&  48.44  $\pm$   2.54&  51.48  $\pm$   3.34&  32.11  $\pm$   4.38&  12.72  $\pm$  0.09  &  31.14  $\pm$  0.42  \\
 LHN20   &   1.64  &   6.48  $\pm$   2.86&  13.90  $\pm$   4.39&  13.16  $\pm$   1.63&   5.65  $\pm$   1.68&   0.00  $\pm$   0.00&  12.35  $\pm$  0.24  &  50.81  $\pm$  4.78  \\
 LHN24   &   2.15  &   2.82  $\pm$   4.05&  21.95  $\pm$   6.87&  16.57  $\pm$   1.77&  15.74  $\pm$   2.11&  12.31  $\pm$   3.81&  12.68  $\pm$  0.15  &  38.64  $\pm$  3.64  \\
 LHN25   &   2.18  &   4.55  $\pm$   2.42&  13.87  $\pm$   4.21&  14.60  $\pm$   3.38&   3.42  $\pm$   3.83&   9.82  $\pm$   2.13&  12.56  $\pm$  0.26  &  38.82  $\pm$  3.61  \\
 LHN27   &   2.23  &   5.23  $\pm$   3.44&  18.90  $\pm$   4.68&  15.47  $\pm$   2.44&  10.34  $\pm$   2.39&  12.49 s $\pm$   3.57&  12.61  $\pm$  0.28  &  52.00  $\pm$  0.68  \\
 LHN29   &   1.96$^{a}$  &   0.00  $\pm$   0.00&  12.83  $\pm$   4.11&  27.47  $\pm$   1.75&  32.45  $\pm$   2.74&  18.45  $\pm$   3.80&  12.51  $\pm$  0.19  &  30.51  $\pm$  1.34  \\
 LHN30   &   1.56  &   9.53  $\pm$   2.82&  22.48  $\pm$   5.10&  20.80  $\pm$   2.00&  11.37  $\pm$   1.62&   0.00  $\pm$   0.00&  12.43  $\pm$  0.18  &  45.75  $\pm$  2.43  \\
 LHN31   &   3.03  &   5.38  $\pm$   2.79&  12.77  $\pm$   4.22&  21.98  $\pm$   1.61&  14.76  $\pm$   2.51&   0.00  $\pm$  0.0&  12.94  $\pm$  0.17  &  62.10  $\pm$  5.87  \\
 GN18  &   2.05  &   -  &  - &   9.46  $\pm$  1.09&  11.72  $\pm$  1.62&   0.00  $\pm$  0.00&  12.38  $\pm$  0.14  &  30.46  $\pm$  2.44  \\
  GN32  &   1.83  &   -  &  - &  41.36  $\pm$  1.02&  41.46  $\pm$  1.62&   0.00  $\pm$  0.00&  12.72  $\pm$  0.14  &  35.70  $\pm$  2.86  \\
  GN34  &   1.83  &   -  &  - &  15.12  $\pm$  1.37&  13.64  $\pm$  2.10&   0.00  $\pm$  0.00&  12.28  $\pm$  0.14  &  35.44  $\pm$  2.83  \\
  GN35  &   1.83  &   -  &  - &  37.47  $\pm$  2.04&  32.22  $\pm$  4.45&   0.00  $\pm$  0.00&  12.69  $\pm$  0.14  &  39.82  $\pm$  3.19  \\
  GN44 &   1.74  &   -  &  - &  13.51  $\pm$  1.55&  16.18  $\pm$  2.65&   0.00  $\pm$  0.00&  12.24  $\pm$  0.13  &  34.46  $\pm$  2.76  \\
  GN46  &   1.66  &   -  &  - &  41.45  $\pm$  1.45&  52.20  $\pm$  2.06&  26.40  $\pm$  4.24&  12.71  $\pm$  0.14  &  36.13  $\pm$  2.89  \\
  GN58  &   1.52  &   -  &  - &  23.59  $\pm$  1.07&  12.59  $\pm$  2.05&   0.00  $\pm$  0.00&  12.41  $\pm$  0.14  &  44.58  $\pm$  3.57  \\

\hline
\end{tabular}
\begin{flushleft}
$^{\rm a}$ IRS spectroscopy by Fiolet et al. (2010)\\
$^{\rm b}$ No available PACS data for GOODS-N in this study
\end{flushleft}
\end{table*}
\clearpage

\begin{table*}
\centering
\caption{Summary of ancillary data.}
\label{catalog}
\renewcommand{\footnoterule}{}  
\begin{tabular}{lcccccccc}
\hline 
ID& RA & DEC &{\it S$_{3.6}$}&{\it S$_{4.5}$} &{\it S$_{5.8}$} &{\it S$_{8.0}$}&{\it S$_{24}$}&{\it S$_{1.4GHz}$}\\
& & &[{\rm $\mu$Jy}] & [{\rm $\mu$Jy}] & [{\rm $\mu$Jy}] & [{\rm $\mu$Jy}] & [{\rm $\mu$Jy}] & [{\rm $\mu$Jy}]\\
\hline
 LHN0  &    161.127548  &     58.921799  &   41.3  $\pm$    0.8&   49.5  $\pm$    1.1&   55.3  $\pm$    4.0&   53.9  $\pm$    3.8&      781  $\pm$   24.0&   101.7  $\pm$   14.2\\
 LHN1  &    161.487091  &     58.888611  &   26.4  $\pm$    0.7&   33.6  $\pm$    1.0&   39.5  $\pm$    3.6&   36.9  $\pm$    3.6&      684  $\pm$   24.1&   314.8  $\pm$   19.1\\
 LHN2  &    161.376022  &     58.920658  &   28.4  $\pm$    0.7&   34.2  $\pm$    0.7&   40.2  $\pm$    3.5&   27.1  $\pm$    2.6&      375  $\pm$   22.3&    29.0  $\pm$    7.5\\
 LHN3  &    161.415726  &     58.906940  &   30.9  $\pm$    0.6&   39.5  $\pm$    0.9&   47.2  $\pm$    3.2&   47.8  $\pm$    3.0&      485  $\pm$   24.1&    46.9  $\pm$    4.3\\
 LHN4  &    161.545685  &     58.879189  &   52.9  $\pm$    0.9&   66.9  $\pm$    1.1&   57.0  $\pm$    3.6&   52.1  $\pm$    3.4&      401  $\pm$   25.0&   160.2  $\pm$   10.7\\
 LHN5  &    161.160263  &     59.075150  &   32.3  $\pm$    0.6&   37.8  $\pm$    0.9&   30.0  $\pm$    3.1&   33.5  $\pm$    3.4&      375  $\pm$   23.3&    51.3  $\pm$    9.8\\
 LHN8  &    161.661163  &     58.936852  &   29.5  $\pm$    0.7&   38.8  $\pm$    1.0&   58.2  $\pm$    3.7&   42.0  $\pm$    3.5&      662  $\pm$   23.4&   159.5  $\pm$    9.9\\
 LHN10  &    161.525223  &     59.141270  &   45.4  $\pm$    0.7&   58.0  $\pm$    1.0&   60.2  $\pm$    3.0&   45.7  $\pm$    3.2&      194  $\pm$   25.6&   238.9  $\pm$   16.2\\
 LHN11  &    161.231583  &     59.167912  &   48.9  $\pm$    0.8&   59.0  $\pm$    1.1&   54.4  $\pm$    3.3&   36.0  $\pm$    3.3&      258  $\pm$   24.8&    45.1  $\pm$   10.0\\
 LHN16  &    161.824844  &     59.042171  &   35.8  $\pm$    0.8&   50.0  $\pm$    1.1&   77.4  $\pm$    4.0&   52.1  $\pm$    3.8&      567  $\pm$   26.4&    55.1  $\pm$    5.4\\
 LHN19  &    161.507462  &     59.154690  &   72.1  $\pm$    1.0&   91.3  $\pm$    1.5&   93.5  $\pm$    3.9&   74.0  $\pm$    3.9&     1011  $\pm$   23.1&    82.2  $\pm$    5.7\\
 LHN20  &    161.676163  &     59.069839  &   53.2  $\pm$    0.8&   61.6  $\pm$    1.2&   57.3  $\pm$    3.3&   42.9  $\pm$    3.5&      307  $\pm$   23.9&    70.6  $\pm$    5.0\\
 LHN24  &    161.843964  &     59.019981  &   31.9  $\pm$    0.6&   41.0  $\pm$    1.0&   41.1  $\pm$    3.1&   38.0  $\pm$    3.3&      439  $\pm$   23.3&    73.7  $\pm$    9.4\\
 LHN25  &    161.933746  &     59.106892  &   36.3  $\pm$    0.5&   43.5  $\pm$    0.8&   46.8  $\pm$    2.8&   36.9  $\pm$    3.3&      656  $\pm$   25.7&    54.9  $\pm$   13.7\\
 LHN27  &    161.729813  &     59.191101  &   28.9  $\pm$    0.6&   36.8  $\pm$    0.7&   40.0  $\pm$    3.5&   38.4  $\pm$    3.0&      337  $\pm$   22.5&    36.4  $\pm$   10.5\\
 LHN29  &    161.909683  &     59.169449  &   44.4  $\pm$    0.6&   51.0  $\pm$    0.8&   52.3  $\pm$    3.1&   47.7  $\pm$    3.3&      688  $\pm$   24.0&    69.2  $\pm$    9.3\\
 LHN30  &    161.944641  &     59.257740  &   50.5  $\pm$    0.6&   59.4  $\pm$    0.9&   40.0  $\pm$    2.9&   50.3  $\pm$    3.3&      341  $\pm$   23.4&    87.6  $\pm$   16.8\\
 LHN31  &    161.935730  &     59.210369  &   35.8  $\pm$    0.6&   44.5  $\pm$    0.9&   39.5  $\pm$    3.1&   35.8  $\pm$    3.4&      404  $\pm$   21.8&     0.0  $\pm$    0.0\\
GN18  &    189.262684  &     62.142494  &   11.4  $\pm$    0.1&   13.9  $\pm$    0.1&   10.3  $\pm$    0.5&   13.3  $\pm$    0.6&      230  $\pm$    7.2&-  \\
 GN32  &    189.256739  &     62.196195  &   53.9  $\pm$    0.1&   70.1  $\pm$    0.1&    8.8  $\pm$    0.3&   57.9  $\pm$    0.4&      716  $\pm$    8.1& - \\
 GN34  &    189.399541  &     62.345261  &   29.3  $\pm$    0.1&   37.5  $\pm$    0.1&   25.3  $\pm$    0.5&   28.2  $\pm$    0.5&      178  $\pm$    5.4&  -\\
 GN35  &    189.076657  &     62.264067  &   14.7  $\pm$    0.1&   19.6  $\pm$    0.1&   22.2  $\pm$    0.4&   20.4  $\pm$    0.4&      314  $\pm$    5.3&  -\\
 GN44  &    189.074144  &     62.235591  &   49.2  $\pm$    0.1&   55.6  $\pm$    0.1&   35.9  $\pm$    0.4&   40.2  $\pm$    0.5&      428  $\pm$    7.2&  -\\
 GN46  &    189.297273  &     62.225206  &   37.9  $\pm$    0.1&   45.0  $\pm$    0.1&   14.8  $\pm$    0.3&   37.8  $\pm$    0.4&      534  $\pm$    8.6&  -\\
 GN58  &    189.294242  &     62.376245  &   38.7  $\pm$    0.1&   47.9  $\pm$    0.1&   12.4  $\pm$    0.7&   46.4  $\pm$    0.6&      383  $\pm$    4.6&  -\\

\hline
\end{tabular}
\begin{flushleft}
\end{flushleft}
\end{table*}
\clearpage

\begin{table*}
\centering
\caption{SHARC-350$\mu$m and MAMBO 1.2mm flux denities of our LHN sample}
\label{catalog}
\renewcommand{\footnoterule}{}  
\begin{tabular}{lcc}
\hline 
ID&{\it S(350$\mu$m)$^{a}$}&{\it S(1.2mm)$^{b}$}\\

 & [{\rm mJy}] & [{\rm mJy}] \\
 LHN1& 39.7$\pm$5.9& 3.08$\pm$0.58\\ 
 LHN8& 31.9$\pm$4.9&2.13$\pm$0.71\\
 LHN16&49.7$\pm$6.5&2.66$\pm$0.78\\
 LHN29&31.8$\pm$5.9&2.48$\pm$0.74\\
\hline
\end{tabular}
\begin{flushleft}
$^{\rm a}$ Flux densities by Kovacs et al. (2010)\\
$^{\rm b}$ Flux demsities by Fiolet et al. 2009
\end{flushleft}
\end{table*}

\end{document}